\documentclass[conference, a4paper]{IEEEtran}
\IEEEoverridecommandlockouts
\usepackage{cite}
\usepackage{url}
\usepackage{amsmath,amssymb,amsfonts}
\usepackage{algorithmic}
\usepackage{graphicx}
\usepackage{textcomp}
\usepackage{xcolor}
\def\BibTeX{{\rm B\kern-.05em{\sc i\kern-.025em b}\kern-.08em
    T\kern-.1667em\lower.7ex\hbox{E}\kern-.125emX}}

\usepackage[pass,a4paper]{geometry}
\begin{document}

\title{PhishOut: Effective Phishing Detection Using Selected Features\\
}
\author{\IEEEauthorblockN{Suhail Paliath}
\IEEEauthorblockA{\textit{College of Technological Innovation} \\
	\textit{Zayed University}\\
	Abu Dhabi, UAE \\
	m80007167@zu.ac.ae}
\and
\IEEEauthorblockN{Mohammad Abu Qbeitah}
\IEEEauthorblockA{\textit{College of Technological Innovation} \\
\textit{Zayed University}\\
Abu Dhabi, UAE \\
m80007215@zu.ac.ae}
\and
\IEEEauthorblockN{Monther Aldwairi}
\IEEEauthorblockA{\textit{College of Technological Innovation} \\
	\textit{Zayed University}\\
	Abu Dhabi, UAE \\
	monther.aldwairi@zu.ac.ae}
}

\maketitle

\begin{abstract}
Phishing emails are the first step for many of today's attacks. They come with a simple hyperlink, request for action or a full replica of an existing service or website. The goal is generally to trick the user to voluntarily give away his sensitive information such as login credentials. Many approaches and applications have been proposed and developed to catch and filter phishing emails. However, the problem still lacks a complete and comprehensive solution. In this paper, we apply knowledge discovery principles from data cleansing, integration, selection, aggregation, data mining to knowledge extraction. We study the feature effectiveness based on Information Gain and contribute two new features to the literature. We compare six machine-learning approaches to detect phishing based on a small number of carefully chosen features. We calculate false positives, false negatives, mean absolute error, recall, precision and F-measure and achieve very low false positive and negative rates. Na{\"\i}ve Bayes has the least true positives rate and overall Neural Networks holds the most promise for accurate phishing detection with accuracy of 99.4\%.
\end{abstract}

\begin{IEEEkeywords}
phishing email, phishing detection, machine learning, features selection
\end{IEEEkeywords}

\section{Introduction}
\label{sec:introduction}
Phishing is a common type of attack to the extent that almost every one of us receives several phishing emails a week \cite{aldwairi2011malurls}. Phishing is an email based attack where the attacker sends an email claiming to be from a legitimate source. The phisher masquerades as a legitimate organization or figure of authority by sending fake emails and requesting urgent response or action. The ultimate goal is to lure the victims to give away private or valuable information that could be used for stealing personal data, identity theft and/or monetary gain \cite{KumarRRS13}. Phishing is a form of social engineering attacks that use human good nature against them. Humans tend to act urgently on a request from their helpdesk asking to click a link to change their password or follow certain instructions to install a new program. Attackers use social engineering techniques and technical subterfuge to steal identity, account credentials, and exploit both individual and corporate networks on the Internet. \\

Phishing has been the main vehicle for delivery of some of the most serious attacks \cite{aldwairi2017GPUs}. A very recent attack against banks and financial institutions around the world that is claimed to have costed billions of dollars in stolen money used a phishing email as a starting point. Kaspersky reported that Carbanak malware targeted bank employees with a specifically designed spear phishing email to eventually install the backdoor \cite{Carbanak}. According to the Anti-Phishing Working Group (APWG) \cite{APWG} around 630,494 unique phishing sites were detected and 3,774 unique brands were targeted by phishing campaigns.

It is very easy to confuse phishing emails with legitimate requests. As a matter of fact, the attackers go out of their way to make their emails look legitimate by including logos and believable wording. However, phishing emails do not all look alike, because effective phishing emails are custom designed for their intended targets. There are many variations and types of phishing emails reported in the literature \cite{Qbeitah2018}. Researchers have used catchy names, such as: 
\begin{itemize}
	\item Pharming: using Domain Name System (DNS) spoofing to redirect the user to a fake website,
	\item Spear phishing: using information available to the attackers to deliver a more individualized approach targeting specific users,
	\item Whaling: going after wealthy individuals,
	\item Vishing: that is voice phishing, where the attacker calls with a recorded message to trick the user to dial in his private information \cite{Hong}.
\end{itemize}

\begin{figure}
	\includegraphics[width=1.0 \linewidth]{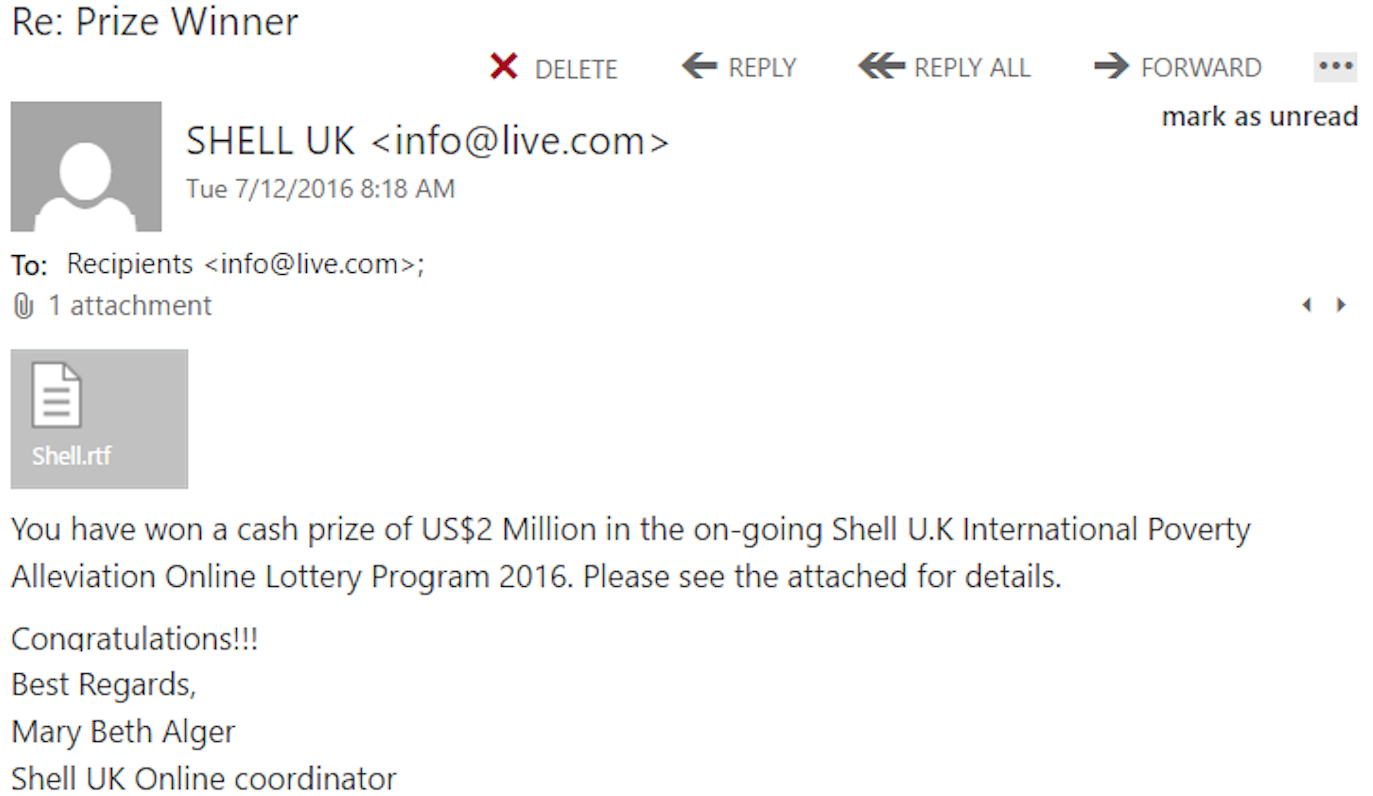}
	\caption{Easy to spot sample phishing email}
	\label{fig:1}       
\end{figure}

\begin{figure}[ht]
	\includegraphics[width=1.0 \linewidth]{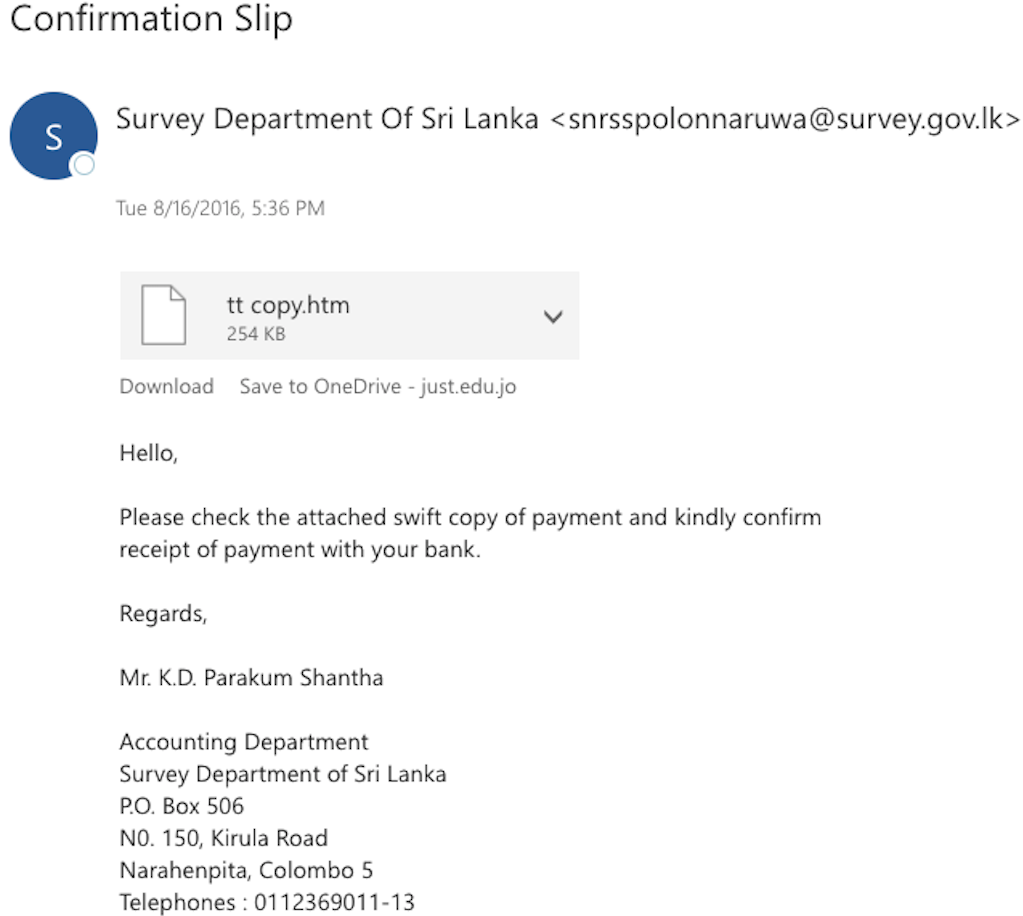}
	\caption{Phishing email close to legitimate emails}
	\label{fig:2}       
\end{figure}

Phishing emails have evolved and mutated into various forms, some are easy to spot such as {fig:1} and others such as Figure \ref{fig:2} are well crafted to look legitimate. The phishing email in Figure \ref{fig:1} preys on human greed, which might blind the recipient to the fact that he/she might have never participated or even heard of such lottery, let alone winning a US\$2 million prize. However, human greed for easy money tends to lure the recipient to open the (.rtf) attachment that often contains a macro, which will kick start the attack. The second phishing email in Figure \ref{fig:2}, is much better crafted to look like a harmless transaction, however the attachment will probably download and execute some sort of malware \cite{Ibrahim2019}.
\\
We cannot but help notice the effort the attackers put to make the emails look benign and the redirection websites look trustworthy. The emails are often sent from accounts that look legit or that belong to a previous business associate or a friend. The requests are well written to look believable, nonetheless, you can notice a few identifying or suspicious features such as: urgent request, existence of a hyperlink or attachment, inaccuracy of the content, and in some instances the sender email is out right suspicious \cite{Masri}. Despite the obvious features of all the previous emails, the majority of average users with low security awareness are susceptible to falling for such schemes. Finally, phishing detection is a complex process, which involves many factors and the parameters that might be vague \cite{aldwairi2012baeza}. Hence, we argue that there is need for a better approach towards phishing detection.
\\
In this paper we study, compare and experiment with several phishing identifying features. In addition, we compare several top machine learning algorithms using real phishing emails taken from a publicly available dataset \cite{Nazario}. The rest of this article is organized as follows. Section \ref{sec:related} surveys the related work in the literature. Section \ref{sec:algorithm} presents the features selection, testing and metrics. Section \ref{sec:analysis} explains the experimental setup and discusses the results. Finally, section \ref{sec:conclusions} lays out the conclusions and directions for future work.

%

%

%

\section{Background and related work}
\label{sec:related}
This section explains the various forms of phishing emails, email delivery process, phishing detection toolbars and the most recent machine learning phishing detection techniques.

\subsection{Phishing forms}
\label{sec:forms}
Based on the techniques used to deliver attacks or steal personal information, we divide the phishing attacks into the following groups \cite{Email.cz}.

\begin{enumerate}
	\item $First$ $group$ is link-based.
	
	\item $Second$ $group$ is text-based, it relies only on appealing to the recipients psyche in order to lure him to give away the password. 
	\item $Third$ $group$ is image-based. The detection of this group is more difficult because it combines images with text as well as a link to the phishing website. 
	\item $Fourth$ $group$ is attachment-based. This category contains malicious software in the form of benign looking HTML files, documents or PDF attachments.
\end{enumerate}




\subsection{Recent works}
\label{sec:Recent works}

Phishing detection toolbars are browser extensions that help users detect phished websites such as Netcraft, SpoofGuard, and CallingID \cite{Toolbars}. This is a client side tool for phishing detection. These methods might be rendered ineffective by spoofing of the DNS cache entries and phishing website can be purported as a legitimate one \cite{AbuNimeh}. Another client side technique is by the use of black and white lists, which are good methods if the attack domains are fixed, and do not change often. However, this is not true as Khonji et al. had stated that 63\% of phishing campaigns last only for two hours \cite{Khonji}. Therefore, we conclude that toolbars or blacklists would not be effective against zero-day phishing attacks.

Machine Learning is a subdomain of Artificial Intelligence, which uses data mining techniques and classification algorithms. In classification type of problems we would train the classifier with a training dataset to learn several features of the input (variables or attributes) and associate those to the appropriate class. Then the classifier is fed a new unclassified dataset and it uses the learnt model to predict which class a new record belongs to, that is, to classify an email as phishing or not \cite{Yaseen2017}. \\
Chandreshekar \cite{Chandrasekaran} used a dataset of 400 emails of which 200 emails were phishing emails and a total of 25 features that included a mixture of style markers and structural attributes were extracted and classified using a one-class Support Vector Machine (SVM). It can be seen that in the five runs of the experiment the accuracy varied largely. That is because words that are seen in phishing emails are added to the feature set and that removing structural attributes from the training set decreased accuracy by 20\%. Moreover, the experiment base used in the test is not large enough to draw a conclusion.\\
Abu-Nimeh \cite{AbuNimeh2} compared six classifiers accuracy 
and used the phishing2.mbox provided by Nazario \cite{Nazario}, where 43 features used to train and test the classifiers. The training and testing included 10-fold cross validations, which are averaged. They concluded that it's difficult to select one classifier and rather it is better to choose a classifier that best fits the research problem. It was found that RF produces the best result with F-Measure 90.24\%, but at the same time, it has the highest false positive rate of 8.29\%. This is not appealing to users, as they would not want a legitimate email to be classified as Spam or phishing. In addition, LR showed high precision of 95\%, but also the highest number of false negatives of 17.04\%. This error might be due to the large number of features and could be optimized by identifying the information gain per feature variable.
\\
Jabri and Ibrahim focused their attention on detecting phishing websites using PRISM algorithm and induction rules. With 20 features and 1,000 test-cases they reported a modest 87\% accuracy with 0.1\% error rate \cite{Jabri}.
For a comprehensive study and analysis of phished website detection techniques the reader is referred to \cite{Varshney}
\\
Mbah, Lashkari and Ghorbani \cite{Mbah} proposed Phishing Alerting System (PHAS) to detect advertisement and pornographic phishing emails. They used WEKA and two classification algorithms: KNN and Decision Tree (J48). KNN achieved best precision and recall of 93.4\% 93.1\%, respectively.
\\
Smadi et al. \cite{Smadi} 
experimented with several classifiers to determine the best algorithm for phsihing emails detection. They achieved the highest accuracy, to date, of 98.87\% with Random Forest algorithm and using 23 features. \\
Finally and more recently, Abutair \cite{Abutair} used Case-Based Reasoning (CBR) to detect phishing emails. They focused on lightweight training with small datasets and limited number of features. Their best reported accuracy was 95.62\%. 
\\We believe their is still more room for improvement in terms of accuracy, especially with using smaller number of features and small datasets to reduce processing times.

\section{Classification approach}
\label{sec:algorithm}

Previous phishing detection techniques focused mostly on classification and prediction. We benefit from Knowledge Discovery process and apply data cleansing, integration, selection, aggregation, data mining, pattern matching to better extract knowledge. Our approach focuses mainly on content-based feature extraction simply because it is simple and proven to be highly effective in phishing detection. We use the public available Nazario phishing corpus \cite{Nazario} as phishing dataset and Spamassasins 20021010\_easy\_ham.tar.bz2 corpus \cite{SpamAssassin} as the ham dataset.

\subsection{Feature selection}
\label{sec:features}
Mbah, Lashkari and Ghorbani \cite{Mbah} surveyed the 20 most commonly used features in the literature. IP URL was the most frequently used feature, followed by dots number in URL, hostname length and the existence of @ symbol. We experiment with the top 20 as well as other feature we handpicked by carefully observing and studying the features in phishing dataset. We used Thunderbird email client to open the emails and found that almost all the phishing emails contain a URL to a malicious website. Most of the times URL obfuscation methods were used to masquerade as a legitimate website. \\
Based on calculating the Information Gain (IG), we ranked all studied features. The following are the features we have selected and used in our classifiers according to the IG scores shown by Table \ref{tab:1}. The features fall into either body or URL features.

\begin{enumerate}
	\item Caps ratio (capRatio). The ratio of uppercase and lowercase letters.
	\item Number of links (NoLinks) in the email \cite{Fette}.
	\item Number of links that have IP addresses (NoLinksIP) as domain name \cite{Chen}.
	\item Number of links with the presence of ASCII characters (NoLinkASCII) \cite{Abawajy}.
	\item Presence of @ in the link (isAtPresent) \cite{Toolan}.
	\item Total number of words in email message (NoWords) \cite{Almomani}.
	\item Number of Phishy words (NoPhishyWords). Words that appear with high frequency in phishing emails, such as money, bank, update, and verify, etc.
	\item Href mismatch(NoLinkMismatch). Most common URL obfuscation techniques where the visible link text is different than the target such as $<$a$>$href="real link"$>$visible link$<$/a$>$ \cite{Chandrasekaran}.
\end{enumerate}	

By using relatively few but prominent features, we were able to detect phishing emails with 99.4\% accuracy (see Section \ref{sec:analysis}). We also preferred the use of quantifying features than binary as these features might hold more information. We contribute two new features: capRatio and NoPhishyWords. According to IG Table \ref{tab:1}, both features are very effective for phishing detection.
\begin{table}
	\caption{Selected feature ranked by Information Gain}
	\label{tab:1}       
	\begin{center}
		\begin{tabular}{lll}
			\hline\noalign{\smallskip}
			Rank & Feature & IG  \\
			\noalign{\smallskip}\hline\noalign{\smallskip}
			1 & NoLinks & 0.8490 \\
			2 & capRatio & 0.3923 \\
			3 & NoLinksIP & 0.2904 \\
			4 & NoWords & 0.2808 \\
			5 & NoLinkMismatch & 0.2743 \\
			6 & NoPhishyWords & 0.0674 \\
			7 & isAtPresent & 0.0580 \\
			8 & NoLinkASCII & 0.378 \\
			\noalign{\smallskip}\hline
		\end{tabular}
	\end{center}
\end{table}

\subsection{Testing and evaluation metrics}
\label{sec:metrics}

For 
evaluation, we chose the commonly used metrics, which are precision, recall, F-Measure and accuracy. In order to compute these metrics, we need to define some terms that make up the Confusion Matrix.
\begin{itemize}
	\item True Positive (TP): Phishing email correctly classified as a phishing.
	\item False Positive (FP): Ham email misclassified as a phishing.
	\item False Negative (FN): Phishing email misclassified as ham.
	\item True Negative (TN): Ham email correctly classified as ham.
\end{itemize}






\subsection{Classification algorithms}
\label{sec:classification}

We compare six well-known classification algorithms included in WEKA (Waikato Environment for Knowledge Analysis). The classifiers include: Neural Networks \cite{NN}, Rough Set Theory \cite{RS}, Na{\"\i}ve Bayes, Support Vector Machines \cite{SVM}, Random Forest \cite{RF} and Random Tree \cite{RT}.

\section{Results and analysis}
\label{sec:analysis}
In this section we explain how the features are extracted, how the experiments were conducted and platform used. The section goes on to present a thorough analysis of the experimental results.

\subsection{Experimental setup}
\label{sec:setup}

The selected features were programmed in C\# to create a ".arff" WEKA attribute file, which can be used as input for the classification. The experiments were performed using WEKA (Waikato Environment for Knowledge Analysis). For our experiments, we used publicly available pre-classified phishing datasets from Nazario dataset \cite{Nazario}. It includes 414 phishing messages. For the Ham we used 842 messages from Apache SpamAssassin \cite{SpamAssassin}. Experiments were conducted with six different types of classification algorithms \{Neural network (NN),Rough set (RS), The Na{\"\i}ve Bayes (NB), SVM, Random Forest (RF) and Random Trees (RT)\} to identify which method performs the best.

The experiments use the eight attributes discussed earlier and perform 10-fold-cross-validation. Cross validation approximates unbiased error estimates by dividing the data into k-subsets. One subset is used for testing and the remaining $k-1$ are used for training. This process is repeated using each of the $k$ subsets as test data and the results are averaged \cite{AbuNimeh}.

\subsection{Experimental results}
\label{sec:experiments}

\begin{figure}[ht]
	\includegraphics[width=1.0 \linewidth]{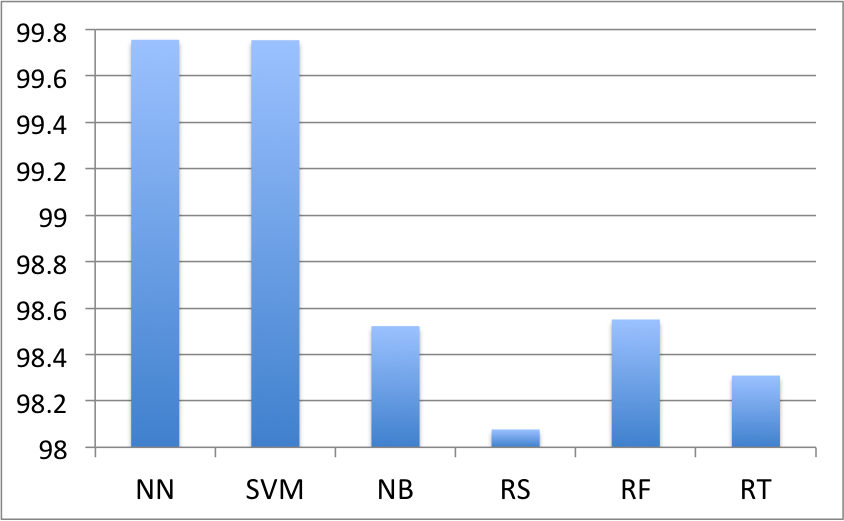}
	\caption{Precision}
	\label{fig:4}       
\end{figure}

Figure \ref{fig:4} shows precision calculated for all algorithms. All algorithms performed very well, however, NN and SVM came out on top with precision of 99.76\% and 99.74\%, respectively. Precision represents positive predictive power, that is how many phishing emails we caught out of all available phishing emails. However, it does not consider our ham prediction capabilities.

Table \ref{tab:2} shows the recall for all algorithms, where NN, RS and RF had the highest recall of 98.55\%. Recall, also referred to as sensitivity or true positive rate, represents the percentage of phishing emails that the filter manages to block. Despite RS and RF high recall, they had the worst and second worst precision, respectively.

F-Measure is shown by Table \ref{tab:2},
which combines the information from the last two measures (precision and recall) through calculating the mean with equal weight for both measures. Neural Networks had the best F-Measure of 99.15\%, which represents NN prediction effectiveness as opposed to the other classifiers.

	\label{fig:5}       


\begin{table}
	\caption{Classifiers Performance (\%)}
	\label{tab:2}       
	\begin{center}
		\begin{tabular}{llllll}
			\hline\noalign{\smallskip}
			Classifier & Precision & Recall & F-Measure & Accuracy & FPR \\
			\noalign{\smallskip}\hline\noalign{\smallskip}
			NN & 99.76 & 98.55 & 99.15 & 99.44 & 0.119\\
			SVM & 99.75 & 97.83  & 98.78 & 99.21 & 0.119\\
			NB & 98.52 & 96.62  & 97.56 & 98.41 & 0.712\\
			RS & 98.08 & 98.55  & 98 .31& 98.89 & 0.949\\
			RF & 98.55 & 98.55  & 98.55 & 99.05 & 0.712\\
			RT & 98.31 & 98.31  & 98.31 & 98.89 & 0.830\\

			\noalign{\smallskip}\hline
		\end{tabular}
	\end{center}
\end{table}

In addition, Table \ref{tab:2} 
shows that NN had the best overall accuracy of 99.44\% among all algorithms. Accuracy is the most accurate measure of prediction power, where it represents the percentage of all correctly classified phishing and ham emails out of all emails. Despite a slight increase in classification time NN has proved to be the best fit for phishing detection.

Moreover, Table \ref{tab:2} shows the False Positives Rate $(FPR = FP / (FP+TN) )$. FPR or fall-out ratio represents the probability of falsely classifying regular emails as phishing or ham emails, which is another measure for accuracy of the classifier. Neural Network as well as SVM exhibited the smallest percentage of FPR at 0.119\%.



Finally, the Mean Absolute Error (MAE) was computed. MAE measures the average magnitude of the errors in a set of forecasts, without considering their class. Support vector machines had the least MAE of 0.8\%, while NN had the second highest MAE of 1.5\%. Nonetheless, all MAE values are relatively small and acceptable given the high TPR.

\section{Conclusions}
\label{sec:conclusions}
 Phishing features from the literature and new ones were proposed. the most effective features according to Information Gain were selected. We contributed two new and effective features: capRatio ranked second among all features with IG of 0.39 and and NoPhishyWords with IG of 0.07. 
 Finally, we compared the predictive capability of various classifiers. We identified Neural Networks to be the best classifier for phishing emails detection with overall accuracy of 99.4\%. However, Neural Networks caused a slight but noticeable degradation in classification performance and had a significant MAE rate of 1.5\%.

Future work includes adding more features such as Document Frequency (DF) and Inverse Document Frequency (IDF) \cite{Balbahaith2017}. We plan to extend our previous work on Multi-classifier system using word embeddings \cite{ABC} to better find unknown phishing forms. 

\section*{Acknowledgment}

This work was supported in part by Zayed University Research Office, Research Cluster Award \# R17079.

\vspace{12pt}
\color{red}
\end{document}